\documentclass[a4paper]{jpconf}
\usepackage{graphicx}
\usepackage{bm}
\begin{document}
\title{Inhomogeneous Pseudogap Phenomenon in the BCS-BEC Crossover Regime of a Trapped Superfluid Fermi Gas}

\author{Ryota Watanabe$^1$, Shunji Tsuchiya$^{2,3,4}$ and Yoji Ohashi$^{1,4}$}

\address{
$^1$ Department of Physics, Faculty of Science and Technology, Keio University, Yokohama, Japan\\
$^2$ Department of Physics, Faculty of Science, Tokyo University of Science, Tokyo, Japan\\
$^3$ Research and Education Center for Natural Sciences, Keio University, Yokohama, Japan\\
$^4$ JST (CREST), Saitama, Japan\\
}

\ead{rwatanab@rk.phys.keio.ac.jp}

\begin{abstract}
We investigate pseudogap phenomena in the unitarity limit of a trapped superfluid Fermi gas. Including effect of strong pairing fluctuations within a $T$-matrix approximation, as well as effects of a harmonic trap within the local density approximation (LDA), we calculate the local superfluid density of states below the superfluid phase transition temperature $T_{\rm c}$. We show that the spatial region where single-particle excitations are dominated by the pseudogap may still exist even below $T_{\rm c}$, due to inhomogeneous pairing fluctuations caused by the trap potential. From the temperature dependence of the pseudogapped density of states, we identify the pseudogap regime of the unitarity Fermi gas with respect to the temperature and spatial position. We also show that the combined $T$-matrix theory with the LDA can quantitatively explain the local pressure which was recently observed in the unitarity limit of a $^6$Li Fermi gas.
\end{abstract}
\section{Introduction}
\par
Ultracold Fermi gases are very useful for the study of many-body physics in strongly interacting fermion systems. Indeed, using a tunable interaction associated with a Feshbach resonance\cite{GIORGINI,BLOCH,OHASHI1}, we can study strong-coupling effects on single-particle properties, as well as thermodynamic properties\cite{GIORGINI,BLOCH,HORIKOSHI,NASCIMBENE}, from the weak-coupling BCS (Bardeen-Cooper-Schrieffer) regime to the strong-coupling BEC (Bose-Einstein condensation) regime in a unified manner. In particular, the so-called pseudogap phenomenon, which is a typical phenomenon in strong-coupling fermion systems, has recently attracted much attention in cold Fermi gases. The photoemission spectra observed in a $^{40}$K Fermi gas\cite{STEWART,GAEBLER} clearly deviates from those for a free Fermi gas in the BCS-BEC crossover region. As the origin of this, the importance of pairing fluctuations has been pointed out\cite{TSUCHIYA,WATANABE,TSUCHIYA2,WATANABE2,HU,HU2}. Since the pseudogap phenomenon has been also discussed in high-$T_{\rm c}$ cuprates\cite{FISCHER}, the study of cold Fermi gases in the BCS-BEC crossover region is expected to be helpful for clarifying various anomalies observed in high-$T_{\rm c}$ cuprates.
\par
In this paper, we investigate the local density of states (LDOS), as well as the local pressure, and effects of pseudogap effects associated with strong pairing fluctuations in a trapped unitarity Fermi superfluid. For this purpose, we employ a combined $T$-matrix theory with the local density approximation (LDA). We note that this strong-coupling theory has succeeded in quantitatively explaining the anomalous behavior of the photoemission spectra observed in a $^{40}$K Fermi gas\cite{TSUCHIYA2,WATANABE2,HU}. Within this theoretical framework, we show that the spatial inhomogeneity due to the trap potential naturally leads to the coexistence of the spatial region where the ordinary BCS excitation gap can be seen and the region where the pseudogap still dominates over single-particle excitations in a gas cloud. We also apply this theory to the local pressure which has been recently observed in a $^6$Li Fermi gas\cite{NASCIMBENE}. For simplicity, we set $\hbar=k_{\rm B}=1$ throughout this paper.
\par

\section{Combined $T$-matrix theory with local density approximation}

We consider a trapped superfluid Fermi gas with two atomic hyperfine states, described by pseudo spin $\sigma=\uparrow,\downarrow$. The model Hamiltonian in the absence of a trap is given by, in the Nambu representation\cite{WATANABE},
\begin{equation}
H=\sum_{\bm p}\Psi_{\bm p}^\dagger[\xi_{\bm p}\tau_3-\Delta\tau_1]\Psi_{\bm p}-\frac{U}{2}\sum_{{\bm q},j=1,2}\rho_j({\bm q})\rho_j(-{\bm q}).
\end{equation}
Here, $\Psi_{\bm p}^\dagger=(c_{{\bm p}\uparrow}^\dagger,c_{-{\bm p}\downarrow})$ is the two-component Nambu field, where $c_{\bm p\sigma}^\dagger$ is the creation operator of a Fermi atom with the kinetic energy $\xi_{\bm p}=p^2/(2m)-\mu$, measured from the chemical potential $\mu$. $\tau_j(j=1,2,3)$ are Pauli matrices acting on the particle-hole space, and the superfluid order parameter $\Delta$ is chosen so as to be proportional to the $\tau_1$ component. $\rho_j=\sum_{\bm p}\Psi_{\bm {p+q}}^\dagger\tau_j\Psi_{\bm p} (j=1,2)$ are the generalized density operators, describing amplitude $(j=1)$ and phase $(j=2)$ fluctuations in the Cooper channel. $-U$ ($<0$) is a pairing interaction which is assumed to be tunable. In this paper, we consider the case of unitarity limit, $a_{\rm s}^{-1}=0$, where the scattering length $a_s$ is related to the pairing interaction $-U$ as $\frac{m}{4\pi a_{\rm s}}=-U+\sum_{\bm p}^{p_c}\frac{1}{2\epsilon_p}$\cite{RANDERIA}.
\par
Effects of a trap is included within the LDA, which is conveniently achieved by replacing the Fermi chemical potential $\mu$ by the LDA expression $\mu(r)\equiv \mu-V({\bm r})$, where $V({\bm r})=m\omega_0^2 r^2/2$ is a harmonic trap. Various quantities then depend on the position $r$. For example, the LDA single-particle Green's function has the form $G_{\bm p}(i\omega_n,r)^{-1}=G_{\bm p}^0(i\omega_n,r)^{-1}-\Sigma_{\bm p}(i\omega_n,r)$, where $G_{\bm p}^0(i\omega_n,r)^{-1}=i\omega_n-\xi_p(r)\tau_3+\Delta(r)\tau_1$ (where $\omega_n$ is the fermion Matsubara frequency), $\xi(r)=p^2/(2m)-\mu(r)$, and $\Delta(r)$ is the LDA superfluid order parameter. $\Sigma_{\bm p}(i\omega_n,r)$ is an LDA self-energy correction, involving effects of pairing fluctuations within the $T$-matrix approximation, which is given by $\Sigma_{\bm p}(i\omega_n,r)=-T\sum_{{\bm q},\nu_n}\sum_{s,s^\prime=\pm}\Gamma_{\bm q}^{ss^\prime}(i\nu_n,r)\tau_{-s}G_{\bm {p+q}}^0(i\omega_n+i\nu_n,r)\tau_{-s^\prime}$. Here, $\tau_{\pm}=\tau_1\pm\tau_2$, and $\nu_n$ is the boson Matsubara frequency. $\Gamma_{\bm q}^{ss^\prime}(i\nu_n,r)=-U/[1+U\Pi_{\bm q}^{ss^\prime}(i\nu_n,r)]$ is the particle-particle scattering vertex, where $\Pi_{\bm q}^{ss^\prime}(i\nu_n,r)=T\sum_{\bm q,\omega_n}{\rm Tr}\left[\tau_sG_{\bm {p+q}/2}^0(i\omega_n+i\nu_n,r)\tau_{s^\prime}G_{\bm {p-q}/2}^0(i\omega_n,r)\right]$ is a pair correlation function. 

\begin{figure}[h]
\centering
\includegraphics[width=\textwidth]{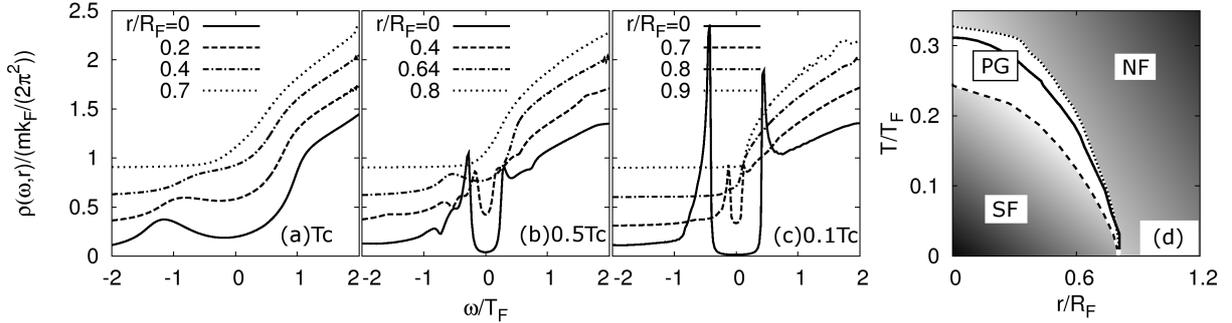}

\caption{Calculated local density of states $\rho(\omega,r)$ in the unitarity limit ($T_{\rm c}=0.312T_{\rm F}$, where $T_{\rm F}$ is the Fermi temperature). (a) $T=T_{\rm c}$. (b) $T=0.5T_{\rm c}$. (c) $T=0.1T_{\rm c}$. In each panel, we have offset the results. $r_{\rm 0}$ is obtained as (a) $0$, (b) $0.64R_{\rm F}$, and $0.80R_{\rm F}$, where $R_{\rm F}=\sqrt{2\varepsilon_{\rm F}/m\omega_0^2}$ is the Thomas-Fermi radius (where $\varepsilon_{\rm F}$ is the Fermi energy).  Panel (d) is the phase diagram in the unitarity limit. ``SF" is the region where the superfluid gap can be seen in LDOS. ``PG" is the region dominated by pseudogap. ``NF" is the region where LDOS is close to that for a normal Fermi gas. The solid line in panel (d) shows $r=r_{\rm 0}(T)$. 
}
\label{FIG1}
\end{figure}

\par
We self-consistently determine the superfluid order parameter $\Delta(r)$ and the chemical potential $\mu$ by solving the LDA gap equation,
\begin{eqnarray}
1&=&U\sum_{\bm p}\frac{1}{2\sqrt{\xi_{\bm p}(r)^2+\Delta(r)^2}}\tanh \frac{\sqrt{\xi_{\bm p}(r)^2+\Delta(r)^2}}{2T},
\label{eq.gap}
\end{eqnarray}
together with the equation for the number $N$ of Fermi atoms,
\begin{eqnarray}
N&=&\int_0^\infty 4\pi r^2drn(r),
\label{BCS}
\end{eqnarray}
where $n(r)=2T\sum_{{\bm p},\omega_n}G_{\bm p}(i\omega_n,r)|_{11}e^{i\omega_n\delta}$ is the particle density in the LDA. The superfluid phase transition temperature $T_{\rm c}$ is determined as the temperature at which the gap equation (\ref{eq.gap}) with $\Delta(r)=0$ is satisfied in the trap center $r=0$\cite{TSUCHIYA2}. We note that, below $T_{\rm c}$, the coupled equations (\ref{eq.gap}) with (\ref{BCS}) give the situation that $\Delta(r)\ne 0$ when $r\ge r_{\rm 0}$ and $\Delta(r)=0$ when $r>r_{\rm 0}$, where $r_{\rm 0}$ becomes large with decreasing the temperature. However, this is an artifact of the LDA, because $\Delta(r)$ should be finite everywhere in the gas below $T_{\rm c}$. Thus, we should regard $r_{\rm 0}$ as a characteristic radius, in the inside of which $|\Delta(r)|$ is large to some extent. 
\par
Once $\Delta(r)$ and $\mu$ are determined, LDOS is calculated from $\rho(\omega,r)=-\frac{1}{\pi}\sum_{\bm p}{\rm Im}G_{\bm p}(i\omega_n\to\omega+i\delta,r)|_{11}$. For the local pressure $P(r)$, using the relation $dP(r)=n(r)d\mu(r)$\cite{NASCIMBENE,HO} for a given temperature, we obtain
\begin{equation}
\label{PRESS}
P(r)=\frac{1}{r}\frac{\partial V(r)}{\partial r}
\int_\infty^rr^\prime dr^\prime n(r^\prime).
\end{equation}

\section{Local Density of States and Local Pressure below $T_{\rm c}$} 
Figure \ref{FIG1}(a) shows the LDOS in unitarity limit at $T_{\rm c}$. Although $\Delta(r)$ vanishes at $T_{\rm c}$, we see a large dip structure around $\omega=0$ in the trap center, originating from strong pairing fluctuations. However, such a dip structure is absent around the edge of the gas. Below $T_{\rm c}$, the pseudogap in the trap center is suppressed by the presence of finite $\Delta(r)$. Instead, the superfluid gap with the coherence peaks at the excitation edges appears (See Figs.\ref{FIG1}(b) and (c)). In panel (b), the pseudogap is found to appear around $r=r_{\rm 0}$. By definition, $1=U\sum_{\bm p}\frac{\tanh\frac{\xi_{\bm p}(r_{\rm 0})}{2T}}{2\xi_{\bm p}(r_{\rm 0})}$ is satisfied at $r=r_{\rm 0}$, which is the same form as the $T_{\rm c}$-equation when $r_{\rm 0}$ is replaced by $r=0$. Thus, as in the case of the trap center at $T_{\rm c}$, strong pairing fluctuations induce the pseudogap around $r=r_{\rm 0}$. 
\par
When we define the pseudogap regime as the region where the dip structure appears in LDOS, we obtain the phase diagram shown in Fig. \ref{FIG1}(d). As expected, the pseudogap regime (PG) exists around the ``$r_{\rm 0}$-line." We briefly note that the pseudogap regime exists even below this line where $\Delta(r)$ is finite. Below this pseudogap regime (denoted by ``SF" in Fig.\ref{FIG1}(d)), the superfluid gap accompanied by the coherence peaks appear in the LDOS. On the other hand, above the pseudogap regime, neither the pseudogap nor the superfluid gap appears in LDOS, which is denoted by ``NF" in panel (d). Figure \ref{FIG1}(d) indicates that the SF region and PG region coexists below $T_{\rm c}$ in a trapped superfluid Fermi gas.
\par
\begin{figure}[t]
\centering
\includegraphics[width=.5\textwidth]{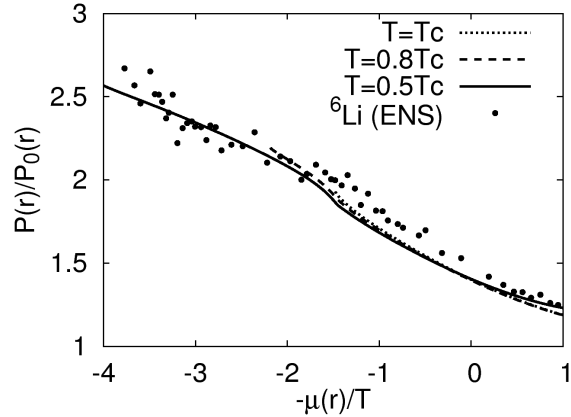}

\caption{Calculated local pressure $P(r)$ in the unitarity limit. Solid circles are experimental results observed by ENS group\cite{NASCIMBENE}. $P(r)$ is normalized by the local pressure $P_0(r)$ in the case of a trapped free Fermi gas with the same chemical potential $\mu$.}
\label{FIG2}
\end{figure}

Strong-coupling effects can be also seen in the local pressure $P(r)$, as shown in Fig. \ref{FIG2}. That is, $P(r)$ is remarkably enhanced from that for a free Fermi gas $P_0(r)$ (where the same chemical potential $\mu$ that used in $P(r)$ is also used.) In this figure, $P(r)$ is found to be almost temperature independent, reflecting the university of the unitarity Fermi gas. As shown in this figure, our results agree well with the recent experiment (solid circles) done by ENS group\cite{NASCIMBENE}.

\section{Summary}
To summarize, we have discussed inhomogeneous pseudogap phenomena in the unitarity limit of a trapped superfluid Fermi gas. Within the framework of the combined $T$-matrix theory with the LDA, we determined the spatial and temperature region in the gas cloud where the pseudogap appears. We also showed how the local pressure is affected by strong pairing interaction. The quantitative agreement of our results with the recent experiment by ENS group indicates the validity of the strong-coupling theory used in this paper for the superfluid Fermi gas in the unitarity regime. Our results would be helpful for the study of strong-coupling effects on trapped Fermi superfluids.

We would like to thank S. Watabe, D. Inotani, and T. Kashimura for fruitful discussions. This work was supported by Global COE Program ``High-Level Global Cooperation for Leading-Edge Platform on Access Spaces (C12)'', as well as the Japan Society for the Promotion of Science. Y. O. was supported by KAKENHI from MEXT in Japan (22540412, 23104723, 23500056).



\section*{References}
\begin{thebibliography}{9}
\bibitem{GIORGINI}S. Giorgini, S. Pitaevskii, and S. Stringari, 2008 Rev. Mod. Phys. \textbf{80} 1215, and references therein
\bibitem{BLOCH}I. Bloch, J. Dalibard, and W. Zwerger, 2008 Rev. Mod. Phys. \textbf{80} 885, and references therein
\bibitem{OHASHI1}Y. Ohashi, and A. Griffin, 2003 Phys. Rev. A \textbf{67} 063612
\bibitem{HORIKOSHI}M. Horikoshi, S. Nakajima, M. Ueda, and T. Mukaiyama, 2010 Science \textbf{327} 442
\bibitem{NASCIMBENE}S. Nascimb\`ene, N. Navon, L. J. Jiang, F. Chevy, and C. Salomon, 2010 Nature \textbf{463} 1057
\bibitem{STEWART}J. T. Stewart, C. A. Regal, and D. S. Jin, 2008 Nature (London) \textbf{454} 744
\bibitem{GAEBLER}J. P. Gaebler, J. T. Stewart, T. E. Drake, D. S. Jin, A. Perali, P. Pieri, and G. C. Strinati, 2010 Nature Phys. \textbf{6} 569
\bibitem{TSUCHIYA}Shunji Tsuchiya, Ryota Watanabe, and Yoji Ohashi, 2009 Phys. Rev. A {\bf 80} 033613 
\bibitem{WATANABE}Ryota Watanabe, Shunji Tsuchiya, and Yoji Ohashi, 2010 Phys. Rev. A {\bf 82} 043630 
\bibitem{TSUCHIYA2}Shunji Tsuchiya, Ryota Watanabe, and Yoji Ohashi, 2010 Phys. Rev. A {\bf 82} 033629
\bibitem{WATANABE2}Ryota Watanabe, Shunji Tsuchiya, and Yoji Ohashi, arXiv:1105.1459. 
\bibitem{HU}Hui Hu, Xia-Ji Liu, Peter D. Drummond, and Hui Dong, 2010 Phys. Rev. Lett. {\bf 104} 240407
\bibitem{HU2}Hui Hu, Xia-Ji Liu, and Peter D. Drummond, 2010 New. J. of Phys. {\bf 12} 063038
\bibitem{FISCHER}O. Fischer, M. Kugler, I. Maggio-Aprile, C. Berthod, and C. Renner, 2007 Rev. Mod. Phys. \textbf{79} 353
\bibitem{RANDERIA}M. Randeria, in \textit{Bose-Einstein Condensation}, edited by A. Griffin, D. W. Snoke, and S. Stringari (Cambridge University Press, New York 1995) p. 355
\bibitem{HO}T. L. Ho and Q. Zhou, 2010 {\it Nature Physics} {\bf 6} 131
\end{thebibliography}
\end{document}